\documentclass{article}

\usepackage[sglblindworkshop, final]{neurips_2025}
\usepackage{fancyhdr}
\usepackage{graphicx}
\usepackage{float}
\usepackage{caption}
\usepackage{subcaption}
\workshoptitle{Generative AI in Finance}

\usepackage[utf8]{inputenc} 
\usepackage[T1]{fontenc}    
\usepackage{hyperref}       
\usepackage{url}            
\usepackage{booktabs}       
\usepackage{amsfonts}       
\usepackage{nicefrac}       
\usepackage{microtype}      
\usepackage{xcolor}         
\usepackage{amsmath}

\title{MASFIN: A Multi-Agent System for Decomposed Financial Reasoning and Forecasting}

\author{%
  Marc S.~Montalvo\thanks{This research was conducted while the author was an undergraduate student at Muhlenberg College.} \\
  Computer Science Department \\
  Rochester Institute of Technology \\
  \texttt{ssm7830@rit.edu}   
  \And
  Hamed Yaghoobian \\
  Department of Math, Computer Science \& Statistics\\
  Muhlenberg College \\
  \texttt{hamedyaghoobian@muhlenberg.edu}
}

\begin{document}




\maketitle
\begin{abstract}
Recent advances in large language models (LLMs) are transforming data-intensive domains, with finance representing a high-stakes environment where transparent and reproducible analysis of heterogeneous signals is essential. Traditional quantitative methods remain vulnerable to survivorship bias, while many AI-driven approaches struggle with signal integration, reproducibility, and computational efficiency. We introduce MASFIN, a modular multi-agent framework that integrates LLMs with structured financial metrics and unstructured news, while embedding explicit bias-mitigation protocols. The system leverages GPT-4.1-nano for reproducability and cost-efficient inference and generates weekly portfolios of 15–30 equities with allocation weights optimized for short-term performance. In an eight-week evaluation, MASFIN delivered a 7.33\% cumulative return, outperforming the S\&P 500, NASDAQ-100, and Dow Jones benchmarks in six of eight weeks, albeit with higher volatility. These findings demonstrate the promise of bias-aware, generative AI frameworks for financial forecasting and highlight opportunities for modular multi-agent design to advance practical, transparent, and reproducible approaches in quantitative finance.
\end{abstract}


\section{Introduction}
Short-term stock prediction is difficult due to volatility, non-stationarity, and the need to integrate quantitative and qualitative signals \citep{schwert1989does}. Recent advances in large language models (LLMs) have renewed interest in this task, with multi-agent frameworks offering modularity, interpretability, and task decomposition \citep{li2025can,yu2024fincon,liu2022finrl}. Yet, existing systems remain limited: many rely on single-modality inputs, focus on either metrics or sentiment, or depend on proprietary data such as BloombergGPT, reducing reproducibility \citep{joshi2025comprehensive,li2025strategy}. Sentiment-driven approaches show promise \citep{mun2025leveraging} but often lack safeguards against common pitfalls in financial research. Traditional quantitative models remain prone to survivorship bias (excluding delisted firms) \citep{brown1992survivorship}, hindsight bias (using future information) \citep{biais2009hindsight}, and overfitting \citep{aliferis2024overfitting}, limiting the transparency and robustness of financial AI.

We introduce MASFIN (\textbf{M}ulti-\textbf{A}gent \textbf{S}ystem for \textbf{Fin}ancial Forecasting), a modular five-stage framework designed to address these shortcomings. MASFIN integrates structured financial metrics from \href{https://finnhub.io/}{Finnhub} and market data from \href{https://finance.yahoo.com/}{Yahoo Finance} with unstructured news sentiment, embedding explicit safeguards against survivorship, hindsight \citep{liu2022finrl}, and overfitting bias. The system is implemented on CrewAI, with agents organized into sequential roles: \textbf{Postmortem} (accounting for delisted firms to prevent survivorship bias), \textbf{Screening} (selecting candidate firms using contemporaneous data to prevent hindsight bias), \textbf{Analysis} (combining financial ratios with sentiment under feature constraints to reduce overfitting), \textbf{Timing} (validating signals with human-in-the-loop oversight), and \textbf{Portfolio} (allocating weights under risk-adjusted constraints). By combining generative reasoning with openly available financial data and bias-aware design, MASFIN provides a transparent, reproducible, and low-cost alternative to proprietary pipelines. Our contributions are:

\begin{enumerate}
    \item \textbf{MASFIN Framework}: We present a five-stage, multi-agent pipeline that integrates Finnhub and Yahoo Finance data with news sentiment and statistical insights, while embedding explicit safeguards against survivorship, hindsight, and overfitting bias. The framework incorporates human-in-the-loop (HITL) validation to mitigate hallucinations, ensures reproducibility with open data and code, and is evaluated in live-market conditions against major benchmarks.
    \item \textbf{Design Principles for Multi-Agent Systems}: Using finance as a high-stakes testbed, we demonstrate how task decomposition, HITL oversight, and bias-aware modular design improve the reliability, interpretability, and affordability of multi-agent generative systems.
\end{enumerate}

By situating finance as a demanding environment, MASFIN contributes a transparent, bias-aware framework that advances the goals of reliability, reproducibility, and robustness in generative AI, with lessons transferable to other high-stakes domains. All implementation details, Python and reproducibility files are available at \href{https://github.com/mmontalvo9/MASFIN}{\texttt{github.com/mmontalvo9/MASFIN}}. The remainder of this paper is organized as follows: Section \ref{sec:system_design} details MASFIN's architecture and methodology, Section \ref{sec:analysis} presents results, Section \ref{sec:limitations} discusses limitations and avenues for future work, and Section \ref{sec:conclusion} concludes.

\section{MASFIN System Design}
\label{sec:system_design}
\begin{figure}[ht]
    \centering
    \vspace{-0.7em}
    \includegraphics[width=\linewidth]{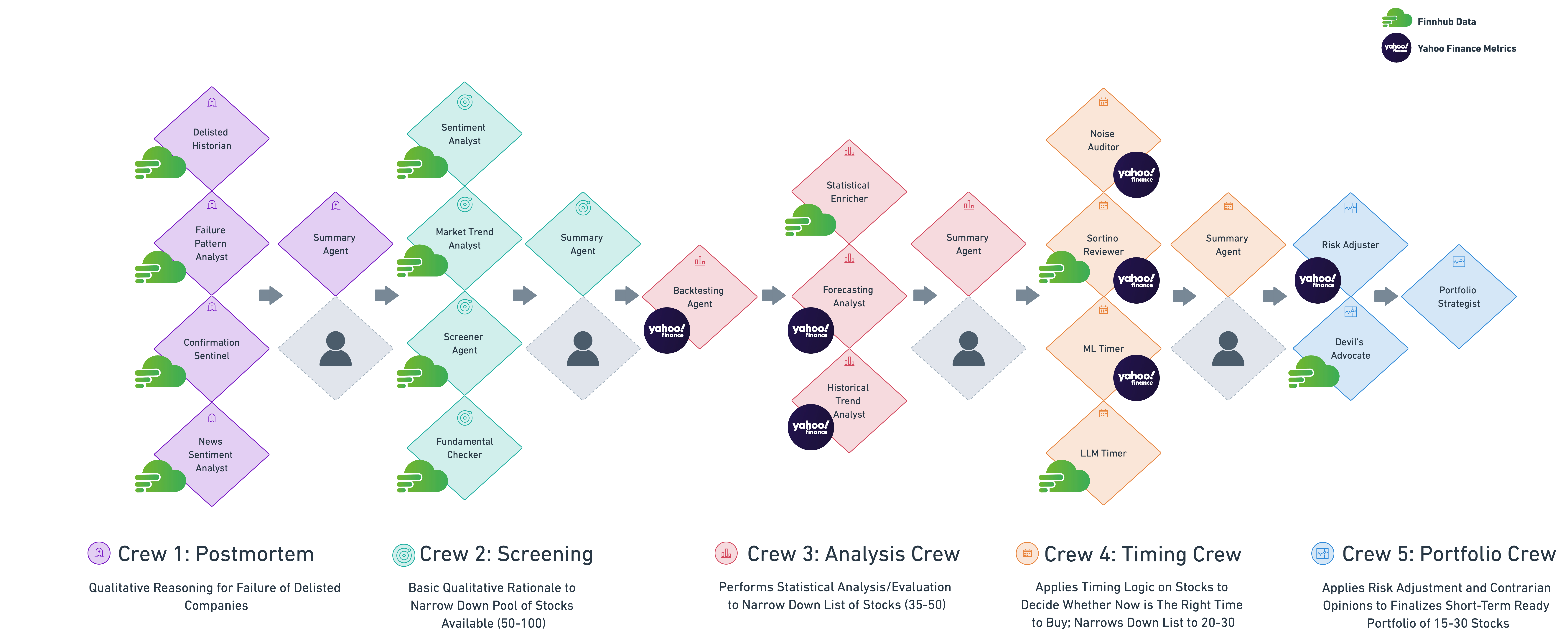}
    \caption{MASFIN's five-stage pipeline with sequential HITL processing.}
    \label{fig:pipeline}
\end{figure}

MASFIN is a multi-agent sequential pipeline for short-term stock forecasting, organized into five crews of 3-5 LLM-based agents. Each contributes to constructing a 15–30 stock portfolio optimized for short-term returns. Except for the Portfolio Crew, all crews include a Summary Agent that synthesizes outputs into a structured report, functioning as the manual handoff in the HITL workflow \citep{buckley2021regulating}. Figure \ref{fig:pipeline} illustrates the pipeline, with agents color-coded by crew, a human figure for manual handoffs, and data sources (Finnhub news and Yahoo Finance) annotated for each agent. 
\vspace{-\baselineskip}

\subsection{Postmortem Crew}
The Postmortem Crew, MASFIN's first stage, identifies risks, failure patterns, and sentiment signals from delisted or at-risk firms while mitigating cognitive and data-driven biases \citep{schropp2024cognitive}. Its inputs are Finnhub headlines from eighteen such companies spanning multiple sectors (Table \ref{tab:delisted-summary}). Agents derive structured features such as historical risk factors, sentiment dynamics, and bias indicators of corporate collapse. By analyzing delisted firms, this stage directly counters survivorship bias, a common omission in financial modeling. A consolidated structured report of failure signals, rationale, mitigation strategies, and references is passed to the Screening Crew as input.

\begin{table}[ht]
\centering
\caption{Examples of delisted or at-risk firms reviewed by the Postmortem Crew.}
\label{tab:delisted-summary}
\resizebox{\linewidth}{!}{
\begin{tabular}{@{}lll c@{}}
\toprule
\textbf{Sector} & \textbf{Ticker} & \textbf{Reason} & \textbf{Date} \\
\midrule
EV / Auto        & NKLA, RIDE, ZEV        & Bankruptcy or listing failure        & 2023–2025 \\
Biotech / Pharma & ADMP, SBBP, CNSP, BLUE & Acquisition, merger, or rule violation & 2021–2024 \\
Consumer / Retail& BBBYQ, REV, GNLN       & Bankruptcy or non-compliance         & 2022–2024 \\
Cannabis         & AGFY, HEXO             & Price or listing failure             & 2023–2024 \\
Tech \& Other    & FRSX, GPRO, SIEB, HYMC & Price or filing failure              & 2020–2025\\
\bottomrule
\end{tabular}}
\end{table}

\subsection{Screening Crew}
The Screening Crew, MASFIN's second stage, filters the market to a shortlist of 50–100 tickers for downstream analysis. Inputs include real-time Finnhub headlines and contextual insights from the Postmortem Crew. Agents perform sentiment evaluation, market trend analysis, and rule-based screening to capture complementary perspectives on candidate stocks \citep{li2025strategy}. A structured, bias-aware shortlist with rationale is passed to the Analysis Crew.

\subsection{Analysis Crew}
The Analysis Crew, MASFIN's third stage, evaluates tickers from the Screening Crew and prior-week holdings with a quantitative framework to identify 35–50 short-term outperformers while minimizing biases, such as hindsight bias \citep{biais2009hindsight}. Inputs include screened tickers, prior-week survivors, weekly snapshots, and percentage changes. Metrics are derived from Yahoo Finance to ensure consistency, while Finnhub headlines provide qualitative context without introducing real-time data leakage. Agents assess indicators such as multi-horizon returns, volatility, Sharpe and Sortino ratios, maximum drawdown, momentum, beta, alpha, return z-scores, volume trends, and moving-average deviations (Table \ref{tab:metrics}). To ensure bias reduction, all analysis metrics are computed using fixed historical windows and contemporaneous data snapshots, preventing look-ahead bias and ensuring temporal alignment across tickers. The crew outputs a validated shortlist enriched with sectoral context and thematic notes, which is then passed to the Timing Crew.

\begin{table}[h]
\centering
\caption{Key MASFIN metrics. The \textit{Global Mean Benchmarking} method contextualizes metrics against the cohort average.}
\label{tab:metrics}
\resizebox{\linewidth}{!}{%
\begin{tabular}{p{0.22\linewidth} p{0.72\linewidth}}
\textbf{Category} & \textbf{Metrics} \\
\midrule
Return-based & 21D Return, 5D Return, Momentum (21D price change) \\
Risk / Adjusted & Volatility (annualized std.), Max Drawdown, Sharpe, Sortino, Beta, Alpha \\
Technical & RSI-14, 5D Return Z-Score, Volume Trend, Price vs. 5D MA \\
Benchmarking & \textit{Global Mean:} $G_m = \tfrac{1}{N}\sum_{i=1}^{N} M_{i,m}$, where $M$ is a metric for ticker $i$ \\
\bottomrule
\end{tabular}%
}
\end{table}

\subsection{Timing Crew}
The Timing Crew, MASFIN's fourth stage, assesses whether candidate tickers from the Analysis Crew are appropriately timed for near-term entry. Inputs include the validated shortlist, Yahoo Finance metrics, and macroeconomic or firm-level context from Finnhub. Agents rely strictly on historical data to prevent hindsight bias and ensure decisions reflect only information available ex-ante \citep{biais2009hindsight,levy2024caution}. Using metrics such as Sortino ratio, return z-score, momentum, regression slope, and trading volume (Table \ref{tab:metrics}), agents generate buy, hold, or sell decisions. A unified decision schema flags inconsistencies or elevated risks, yielding a refined list of 20–30 candidates with justified timing for portfolio entry.

\subsection{Portfolio Crew}
The Portfolio Crew, MASFIN's final stage, consolidates prior outputs into a 15–30 stock portfolio with allocation weights. It resolves conflicts, challenges weak or bias-prone selections, and ensures diversification. Agents use Yahoo Finance for quantitative measures (volatility, Sharpe, Sortino, drawdowns (Table \ref{tab:metrics})) and Finnhub for qualitative validation via sentiment and macro context. The portfolio emphasizes systematic bias control, resisting overfitting and hindsight bias while balancing risk, limiting concentration, and grounding each inclusion in both statistical evidence and external signals \citep{roberts2007exploring,biais2009hindsight,aliferis2024overfitting}.
\vspace{-\baselineskip}

\section{MASFIN Performance Analysis}
\label{sec:analysis}
\begin{figure}[!hbt]
  \centering
  \begin{subfigure}{0.49\linewidth}
    \includegraphics[width=.98\linewidth]{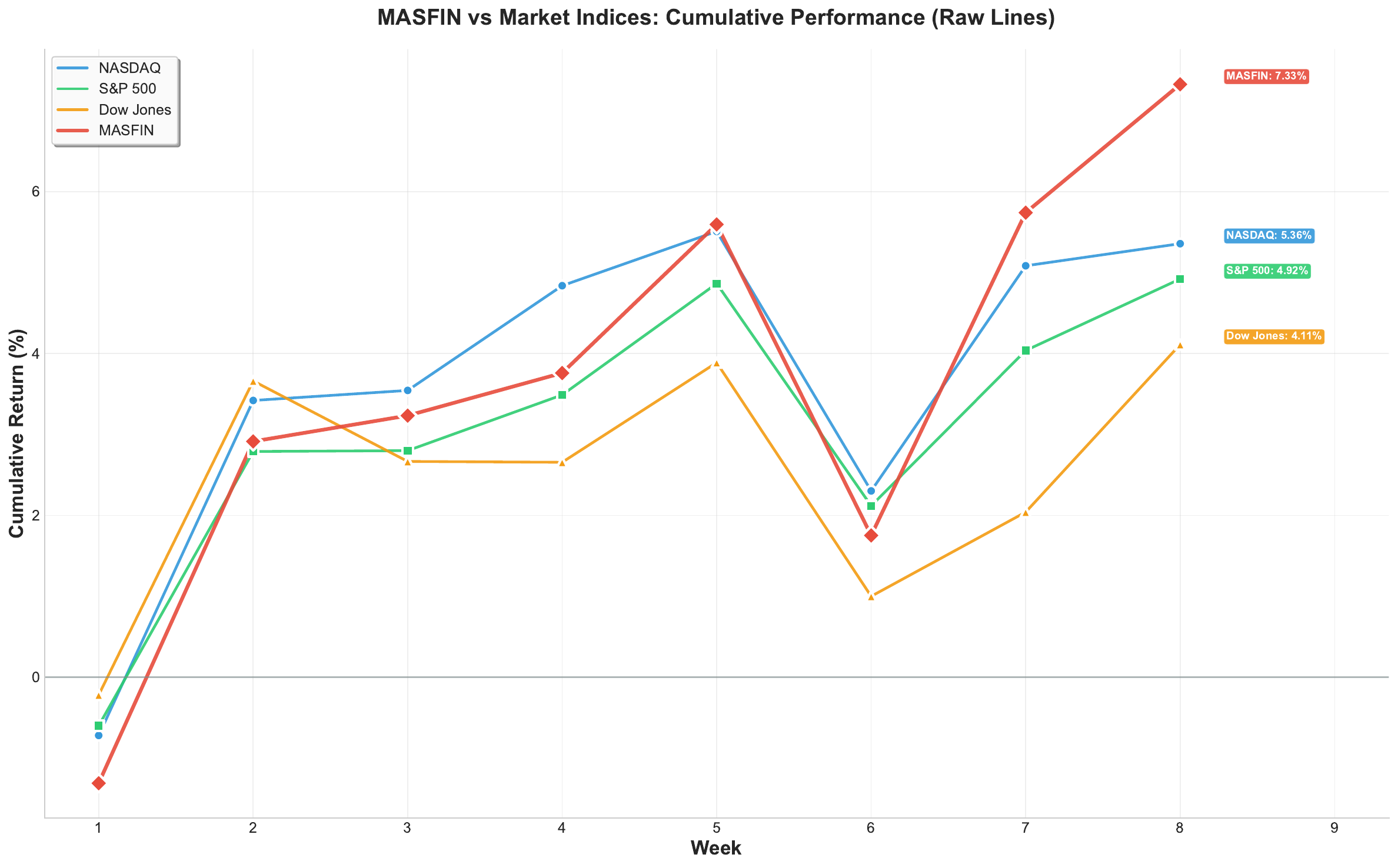}
    \caption{\small Cumulative returns over 8 weeks.}
    \label{fig:growth}
  \end{subfigure}\hfill
  \begin{subfigure}{0.49\linewidth}
    \includegraphics[width=.98\linewidth]{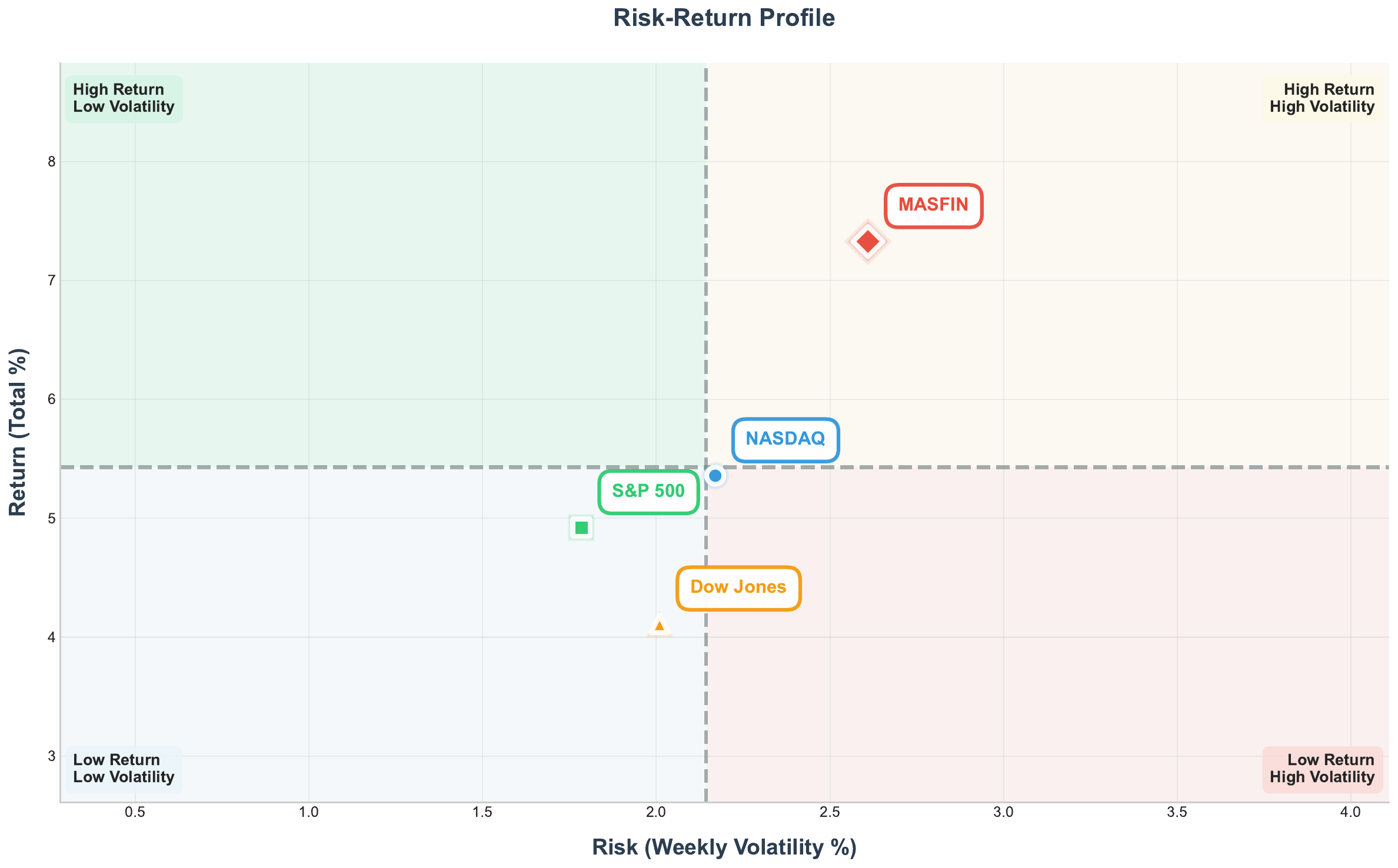}
    \caption{\small Risk–return profile.}
    \label{fig:risk-return}
  \end{subfigure}
  \caption{\small MASFIN performance over an 8-week live evaluation.}
  \label{fig:masfin-performance}
\end{figure}

We evaluated MASFIN over an eight-week horizon with weekly rebalancing, benchmarking against the S\&P 500 (SPY), NASDAQ-100 (QQQ), and Dow Jones Industrial Average (DIA). As shown in Figure \ref{fig:growth}, MASFIN achieved a cumulative return of 7.33\%, surpassing the NASDAQ (5.36\%), S\&P 500 (4.92\%), and Dow Jones (4.11\%). It delivered positive returns in six of eight weeks, a 75\% win rate comparable to the NASDAQ and S\&P 500, and consistently outperformed in total return.

This outperformance was accompanied by higher risk. As shown in Figure \ref{fig:risk-return}, MASFIN’s weekly volatility was 2.61\%, the highest among all benchmarks, placing it in the high-risk, high-return quadrant. Nevertheless, the magnitude of gains suggests a favorable risk-adjusted profile. MASFIN's strong correlations with the S\&P 500 (0.97) and NASDAQ (0.95) suggests amplified performance within existing market trends.



\vspace{-\baselineskip}
\section{Limitations}
\label{sec:limitations}

End-to-end architectures ingesting all metrics, headlines, and analyses proved impractical, exceeding LLM context limits and reducing interpretability \citep{hosseini-etal-2025-efficient}. Fully automated execution also raised our API costs. To address these constraints, we adopted a human-in-the-loop (HITL) workflow with manual handoffs \citep{buckley2021regulating}, enabling validation of outputs, mitigation of hallucinations and bias, and adjustment of agent behavior.

Currently, MASFIN produces weekly portfolios and evaluates them with short-term metrics but lacks a learning mechanism and statistical inference tools such as confidence intervals or hypothesis testing \citep{de2018advances}. Restricting evaluation to eight weeks limits overfitting to specific market regimes while ensuring reproducible testing. These trade-offs emphasize interpretability over automation. Future work includes comparing MASFIN with other AI-based financial systems to better contextualize its performance.

\vspace{-\baselineskip}

\section{Conclusion}
\label{sec:conclusion}
This paper introduced MASFIN, a modular multi-agent system that integrates generative AI with financial metrics and news for short-term portfolio construction under explicit bias-mitigation protocols. In an eight-week evaluation, MASFIN outperformed benchmarks in six of eight weeks, albeit with higher volatility. Its modular design and HITL workflow balanced interpretability and reproducibility against the challenges of context length, computational cost, and lack of formal validation. These results suggest that bias-aware, modular frameworks can enhance the reliability and accessibility of generative AI in high-stakes domains, with financial forecasting serving as a representative testbed for how such systems may evolve \citep{joshi2025comprehensive}.

\section*{Acknowledgments}
This research was supported by the Department of Mathematics, Computer Science, and Statistics and by a Summer Research Grant from the Dean of Academic Life at Muhlenberg College.


\small 
\bibliographystyle{plainnat}
\bibliography{sources}

\end{document}